# Statistical characterization of the yield stress of nanoparticles

Liang Yang[1], Jianjun Bian[2], Weike Yuan[2], Gangfeng Wang[2*]

[1] School of Materials Engineering, Jiangsu University of Technology, Changzhou 213001, PR China

[2] Department of Engineering Mechanics, SVL, Xi'an Jiaotong University, Xi'an 710049, PR China

[*] E-mail address: wanggf@xjtu.edu.cn

**Abstract**

Atomistic simulations are performed to study the statistical mechanical property of gold nanoparticles. It is demonstrated that the yielding behavior of gold nanoparticles is governed by dislocation nucleation around surface steps. Since the nucleation of dislocations is an activated process with the aid of thermal fluctuation, the yield stress at a specific temperature should exhibit a statistical distribution rather than a definite constant value. Molecular dynamics simulations reveal that the yield stress follows a Gaussian distribution at a specific temperature. As the temperature increases, the mean value of yield stress decreases while the width of distribution becomes larger. Based on numerical analysis, the dependence of the mean yield stress on temperature can be well described by a parabolic function. Present study illuminates the statistical features of the yielding behavior of nanostructured elements.

# 1. Introduction

Compared to their macroscopic counterparts, nanostructured elements usually exhibit extraordinary and size-dependent mechanical properties [1]. For example, the experimentally measured moduli of metallic nanowires increase dramatically with decreasing diameters [2][3]. The compressive strength of gold nanopillar is about 800 MPa, close to its ideal shear strength [4]. Experiments reveal that the hardness of silicon nanoparticles is more than three times as that of bulk silicon [5][6]. Among these unusual properties, yield strength is one of the most important mechanical properties of nanomaterials, and great effort has been devoted especially for nanoparticles. For example, gold nanoparticles exhibit higher yield strength than those of bulk gold [7], where particle radius and surface steps together govern the size-dependent yield strength [8]. The strength of CdS nanoparticles extracted from *in-situ* mechanical tests approaches to the idea shear strength of bulk materials [9]. However, different from that at macroscale, the yield strength of nanostructures displays the scattering feature in experiments [10][11], which usually has been attributed to experimental errors or sample nuance. And the maximum value or the mean value with standard deviation is commonly used to quantify the yield strength of nanosized elements.

Owing to the small volume and high surface-to-volume ratio, the initiation of yielding in nanosized elements is closely related to local atoms around surfaces or surface defects, such as for nanowires [14][15][16] and nanoparticles [8][17][18][19]. It is known in statistical thermal dynamics that, for a collection of atoms at a specific temperature, the state of atoms like kinetic energy follows some statistical distribution. Therefore, it is expected that, for a small cluster of atoms like nanostructured elements, the thermal fluctuations of atoms around surfaces may lead to the stochasticity of some mechanical properties like yield stress. Recently, Mordehai *et al.* calculated the yield stress of a faceted Mo nanoparticle at different temperature, and proposed a method to extract the activation free-energy barrier for dislocation nucleation [13]. For nanoparticles, revealing the statistical feature of yield stress is of critical importance to understand their mechanical properties.

In this paper, based on large-scale molecular dynamics simulations, we mainly study the yielding behavior of defect-free gold nanoparticles, and focus on the statistical features of yield stress. Furthermore, the dependence of yield stress on temperature is also explored.

## 2. Computational methods

The uniaxial compressions of gold nanoparticles are performed by using the open-source MD simulation package, LAMMPS [20]. A well-tested embedded atom method (EAM) potential is utilized to describe the atomic interactions among gold atoms, which shows a good agreement with such experimental data as lattice constant, elastic constants, and stacking fault energy [21]. All nanoparticles are carved out of single crystalline bulk gold with a lattice constant of 4.08 Å. The radius of one nanoparticle is denoted by that of its spherical cutting surface. In our simulations, nanoparticles with radii of 8, 12 and 20 nm are investigated. The deformation behaviors at different temperature are studied, from low temperature, 10 K, to room temperature, 300 K. At each temperature, the simulations are repeated 100 times for each sized nanoparticle. Nanoparticles with the same radius keep the identical geometrical profile, but are assigned with various random seeds to initialize the velocities of atoms. Therefore, the initial velocities of the atoms in one nanoparticle follow the Gaussian distribution, while the initial velocity of an individual atom varies in different simulations. Canonical ensembles (NVT) are adopted to describe the atomic systems, and the Nosé–Hoover thermostat is utilized to control the temperature of the system [22][23]. The time integration is implemented based on the velocity-Verlet algorithm with a time step of 0.002 ps.

Uniaxial compression is performed along [111] crystal orientation as depicted in Figure 1. Two rigid planar indenters are adopted to conduct the uniaxial compression of a nanoparticle. The origin of the coordinate system is set at the center of the nanoparticle, with the $z$-axis along [111] orientation. The top and bottom indenters are parallel to {111} facet. A repulsive potential is used to describe the interactions between the rigid indenters and gold atoms, as the same as the setup in our previous study on gold nanoparticles [8]. Prior to compression, the nanoparticle is initially relaxed using conjugate gradient method to achieve a stable configuration with minimum cohesive energy, and then the whole system is dynamically equilibrated at a specific temperature for about 30 ps. Uniaxial compressions are accomplished by moving the two rigid planar indenters simultaneously towards the center of nanoparticle at a constant speed of 0.05 Å/ps.

The compression load $P$ is obtained from the reactive force applied on the indenter, and the compression depth $\delta$ is denoted by the displacement of one indenter. During the simulation, the real contact area $A$ is calculated by Delaunay triangulation algorithm, and the average contact stress $\sigma$ is defined as the compression load divided by the current contact area. To analyze the atomic processes inside the nanoparticle during compression, an open-source package OVITO is used to identify and visualize the characteristics of the nucleated defects [24], and a dislocation extraction algorithm is used to calculate the total dislocation length [25].

## 3. Results and discussions

To precisely quantify the yield point, we first examine the deformation behavior of one gold nanoparticle with radius of 12 nm at temperature 100 K. Owing to the symmetry of the atomic model, only half of the nanoparticle needs to be considered. The loading curve is shown in Figure 2. It clearly shows that the load response can be subdivided into elastic and plastic stages. In the elastic stage, the compression load is linearly accumulated with the increment of the compression depth. When yielding occurs, the continuously building-up of the load is abruptly punctuated, indicating the onset of plasticity. And in the following stage, compression load begins to fluctuate with compression depth, resulting from dislocation activities inside the nanoparticle. The contact area in the elastic stage almost keeps constant, but soon after the yielding point, it abruptly jumps to another constant value and then keeps increasing. Based on the obtained load and the contact area, the average contact stress is calculated as well. After linearly reaching the yield point, the contact stress drops prominently from a local peak to a low stress level.

Atomic surface steps play key roles in the elastic deformation of nanoparticles. In the initial elastic stage of the compression, the planar indenter only touches the out-most surface step, thus the contact area keeps constant, and the load-depth response follows a flat punch model [26]. When yielding occurs, the indenter quickly flattens the out-most surface step and gets into contact with the second one. Simultaneously, partial dislocations nucleate heterogeneously around the fringes of the first surface step. The trailing partial dislocation emits shortly after the nucleation of the leading partial dislocation, forming an extended full dislocation with a stacking fault ribbon. From the yield point, the total length of dislocations $L_{disl}$ increases from zero. Therefore, the local peak of the contact stress right before dislocation nucleation is reasonably defined as the yield stress of a gold nanoparticle.

In order to study the statistical yielding behavior of nanoparticles, the simulation with the 12-nm-radius nanoparticle is repeated 100 times. Due to different random seeds used to initialize atom velocities, although the atom velocities in different simulations follow the Gaussian distribution, the kinetic state of individual atom varies. Since surface dislocation nucleation leads to yielding and the kinetic states of surface atoms are different in various simulations, it is expected that the geometrically identical nanoparticles may display stochasticity in yield stress due to thermal statistical nature. Figure 3a shows the distribution histogram of the yield stress, it follows a Gaussian distribution. And the probability density function of the yield stress can be characterized by

$$\phi(\sigma_Y) = \frac{1}{\sqrt{2\pi}w} \exp\left[-\frac{(\sigma_Y - \sigma_m)^2}{2w^2}\right], \quad (1)$$

where $\sigma_m$ and $w$ are the mean and the standard deviation of the sample data set, respectively. For nanoparticles with radius of 12 nm at 100 K, $\sigma_m$ is 11.67 GPa and $w$ is 0.54 GPa. With respect to the lower, medium and higher yield stress in the statistical distribution, the characteristic dislocations are shown in Figure 3b from left to right. It is noted that, beneath the fringes of the first atomic surface step are the preferential dislocation nucleation sites. On account of the thermal motion of the atoms, the atomic kinetic energy in the nanoparticle demonstrates a statistical distribution. Even for the same nanoparticle at a specific temperature, dislocations may nucleate at different locations, resulting in the variation of yield stress.

Further, we compress the 12-nm-radius nanoparticle at different temperatures, to obtain more information about distribution of yield stress. At each temperature, the simulation is also repeated 100 times, and the distribution histograms of yield stress are displayed in Figure 4b. It is shown that, the yield stress at higher temperature exhibits a wider distribution range, and the mean yield stress decreases gradually with increasing temperature. The atomic energy supplies a thermal assistance to the surface dislocation nucleation and hence to the yielding. At higher temperature, both the average atomic energy and the thermal fluctuation become larger, which leads to the decrease of the mean yield stress and the increase of the distribution width.

To confirm if the statistical distribution of yield stress is an inherent character, nanoparticles with different radii are considered as well. Based on our previous research, atomic surface steps under the indenter play an important role in the yield process and cause great change in yield stress [8]. In order to eliminate the influence of atomic surface steps, we compress nanoparticles with similar surface morphology, therefore, the yield stress of nanoparticles with different radii can be compared directly. Here, nanoparticles with radii of 8 and 20 nm are investigated, the distributions of yield stress at different temperatures are shown Figure 4a and 4c. From these distribution histograms, similar conclusions can be drawn. The yield stress of each set exhibits an approximately Gaussian distribution. As the temperature increases, the mean yield stress decreases. The standard deviation of the yield stress also varies with temperature. To characterize the variation, the coefficient of variation (CV) is calculated as the ratio of the standard deviation to the mean value. As shown in Figure 4d, when the temperature increases from 10 K to 300 K, the CV monotonically increases from ~ 2 % to ~ 7 % for all three sizes of nanoparticles, and the variations of CV for different sized nanoparticles are close to each other.

Now, our simulations demonstrate that, for different sized nanoparticles, no matter at what temperature the simulations are performed, the yield stress is not a constant value, but exhibits a Gaussian distribution. It is reasonable to quantify the yield stress by the mean value with standard deviation. Figure 5 shows the variations of the yield stress of nanoparticles versus the temperature. For all three sizes of nanoparticles, the mean yield stress declines with increasing of temperature. Nevertheless, the trend "smaller is stronger" is satisfied at a given temperature.

To understand the temperature-dependent feature of the mean yield stress, a simple model is utilized. Dislocation nucleation from surface are mainly governed by two factors, the athermal stress and thermally activated stress [12][27]. The former is associated with the elastic limit of material at zero-K temperature, under which dislocation nucleates spontaneously without the assist of thermal fluctuation. The thermally activated stress originates from the thermal vibrations of atoms. Considering these two factors, the yield stress $\sigma_Y$ of one nanoparticle can be given by

$$\sigma_Y = \sigma_a - \sigma_t, \tag{2}$$

where $\sigma_m$ and $\sigma_t$ represent the athermal stress and the thermally activated stress, respectively. The thermally activated stress results in the temperature dependence. In a solid crystal at finite temperature, atoms could be regarded as independent simple harmonic oscillators vibrating around their equilibrium positions. As is known the thermal vibrations of atoms in the nanoparticle will be intensified with rising temperature. The average kinetic energy per-atom is estimated as $E_K = 3k_B T/2$, where, $k_B$ is the Boltzmann constant, $T$ is temperature. Assuming the effective force constant of the metallic bonds is $k_s$, the atomic vibration magnitude would be $\Delta = (2E_K/k_s)^{1/2}$. It should be noted that, the effective force constant of surface atoms may depend on the size of nanoparticles. The atomic thermal vibration induces an atomic-scale strain estimated as $\varepsilon_t = \Delta/a$, where $a$ is the lattice constant. Consequently, the thermally activated stress is estimated as $\sigma_t = E\varepsilon_t$, and $E$ is the material modulus. After substituting to Eq. 2, the relation between temperature and yield stress is expressed as

$$\sigma_Y = \sigma_a - \alpha T^{1/2}, \tag{3}$$

where $\alpha = \frac{E}{a}\left(\frac{6k_B}{k_s}\right)^{1/2}$ is the combination of the above referred parameters. Form Eq. 3, the yield stress shows a parabolic dependence of temperature. In Figure 5, Eq. 3 is used to fit the simulation data, and shows excellent agreement with MD results. For the three nanoparticles with radii of 8 nm, 10 nm, and 12 nm, the fitted athermal yield stresses are 15.43, 13.19, and 10.02 GPa, respectively. In order to estimate the athermal yield stress, we also perform simulation at 0.01 K, and the yield stress for the three nanoparticles are 15.74, 13.85 and 10.57 GPa. The fitted athermal yield stresses by Eq. (3) are close to those obtained directly by MD simulations at 0.01 K, which confirms the applicability of theoretical analysis.

## 4. Conclusion

In this paper, we investigate the statistical characterization of yield stress in the compression of gold nanoparticles. Under uniaxial compression, nanoparticles yield when partial dislocations nucleate from the fringe of surface steps. Since the dislocation nucleation is a thermally activated process, the thermal fluctuation of an atomic system influences the yield stress. At a specified temperature, the yield stress of nanoparticles is not a constant value, but exhibits a Gaussian distribution. As the temperature increases from 10 K to 300 K, the mean value of yield stress decreases while the width of distribution becomes larger. The coefficient of variation increases but keeps the same trend for all three sizes. Based on numerical analysis, a simple model with a parabolic relation is proposed to explain the temperature dependence of yield stress, and shows a good agreement with simulation results. This study is helpful to understand the statistical features of the yielding behavior of nanostructures.

**Acknowledgements**

Support from the National Natural Science Foundation of China (Grant No. 11525209) is acknowledged.

**Figure captions:**

Figure 1. Schematic of the uniaxial compression of gold nanoparticle in MD simulations.

Figure 2. The variations of compression load, contact area, average contact stress and total dislocation length with respect to the accumulation of compression depth. (The vertical line in grey indicates the yield point.)

Figure 3. (a) The distribution of yield stress at 100 K of gold nanoparticles with radius of 12 nm. (b) Characteristic dislocations nucleation beneath surface steps at the onset of plasticity. (Atoms in perfect lattice are not shown for clearness, atoms in red represent surface and dislocation cores, atoms in blue dislocation embryos, and atoms in green stacking faults.)

Figure 4. (a), (b) and (c) The distributions of yield stress at different temperature of three sizes of nanoparticles. (d) The variations of CV versus temperature.

Figure 5. The variations of yield stress with temperature (Symbols) and the fitting lines according to Eq. (3) (Dash-dot lines).

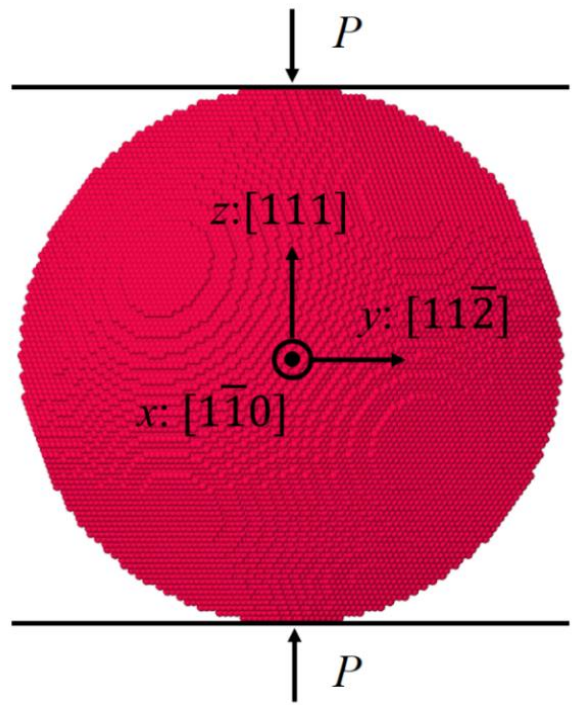

Figure 1. Schematic of the uniaxial compression of gold nanoparticle in MD simulations.

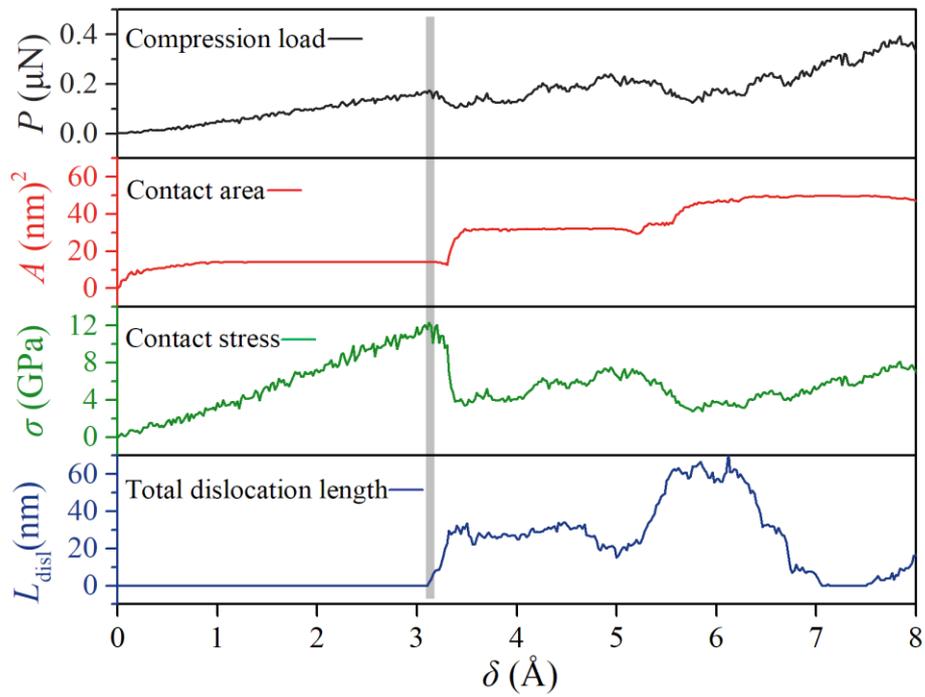

Figure 2. The variations of compression load, contact area, average contact stress and total dislocation length with respect to the accumulation of compression depth. (The vertical line in grey indicates the yield point.)

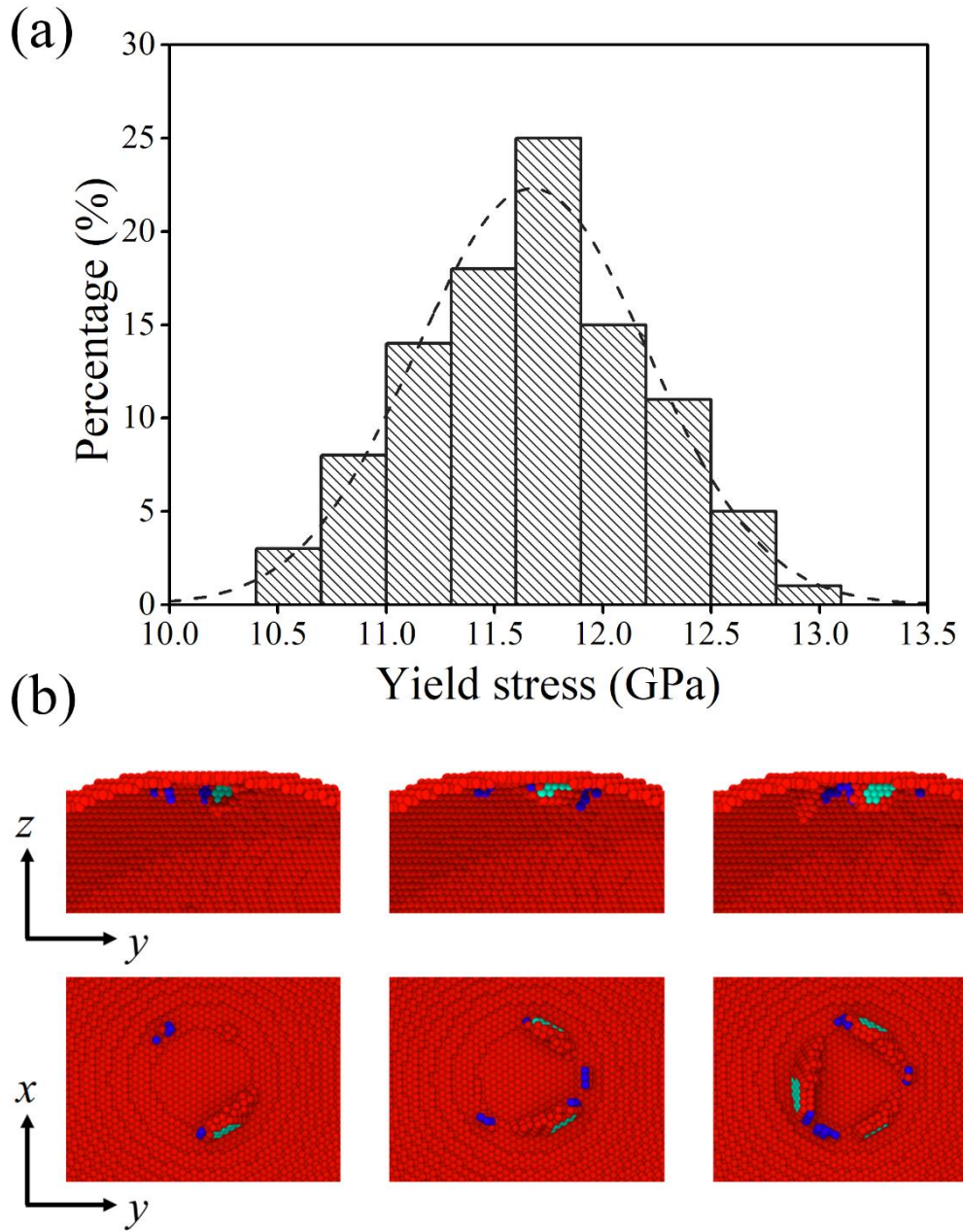

Figure 3. (a) The distribution of yield stress at 100 K of gold nanoparticles with radius of 12 nm. (b) Characteristic dislocations nucleation beneath surface steps at the onset of plasticity. (Atoms in perfect lattice are not shown for clearness, atoms in red represent surface and dislocation cores, atoms in blue dislocation embryos, and atoms in green stacking faults.)

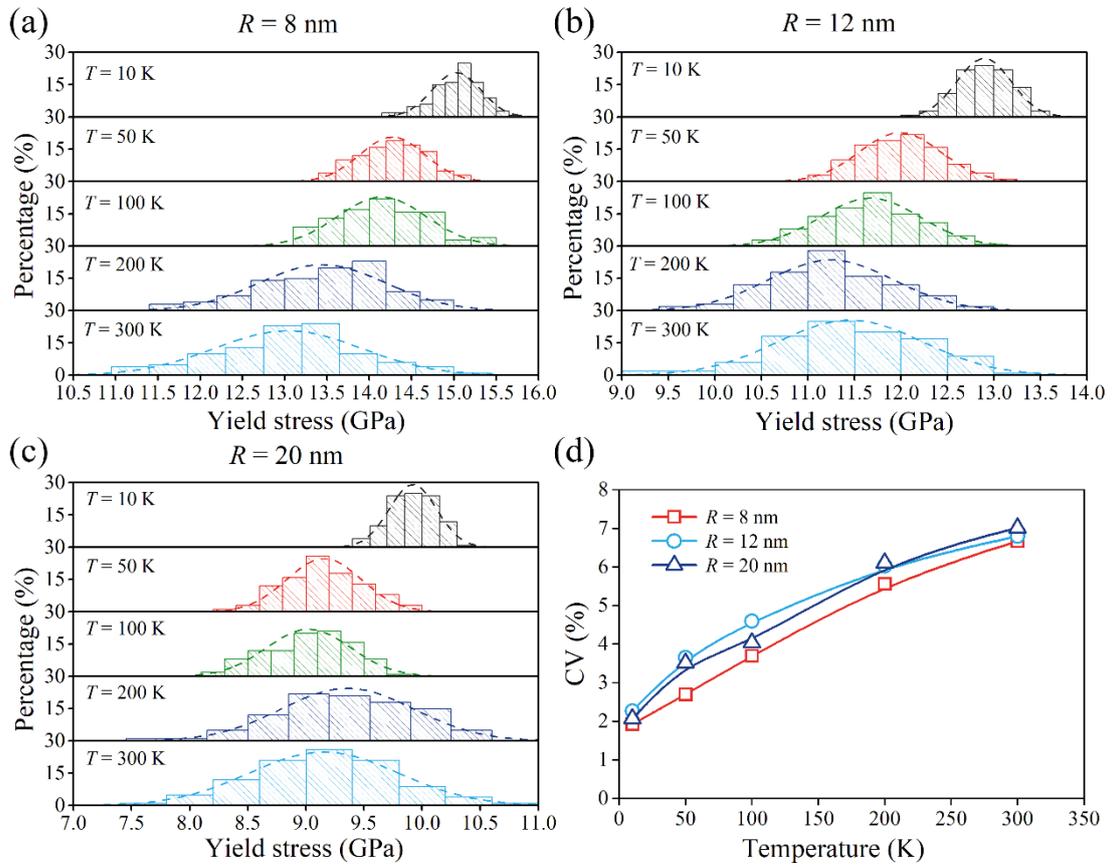

Figure 4. (a), (b) and (c) The distributions of yield stress at different temperature of three sizes of nanoparticles. (d) The variations of CV versus temperature.

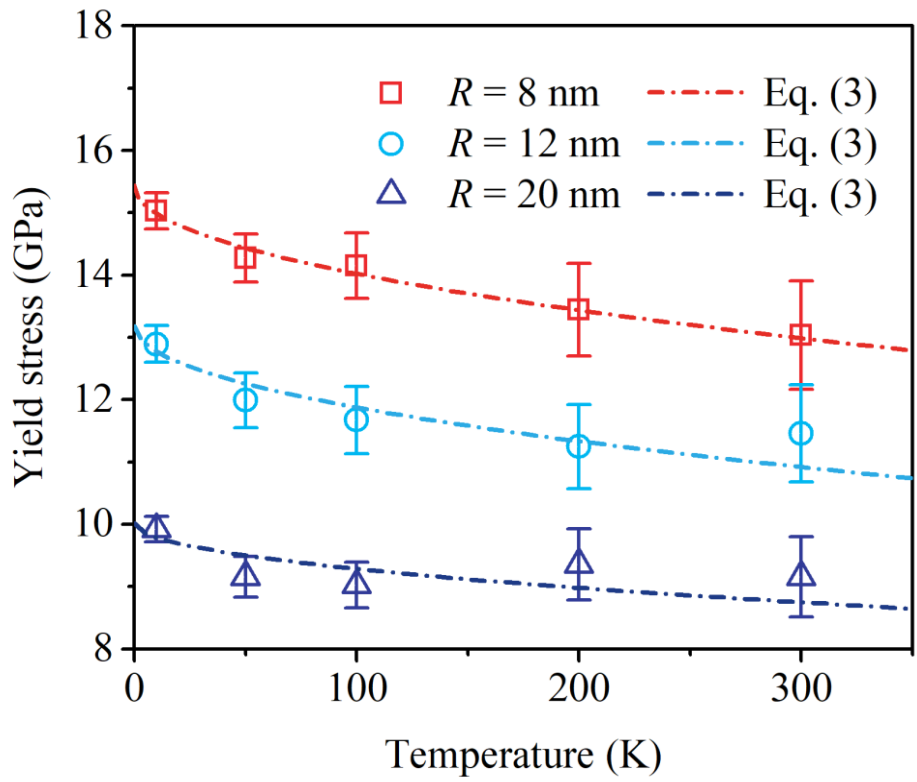

Figure 5. The variations of yield stress with temperature (Symbols) and the fitting lines according to Eq. (3) (Dash-dot lines).